\documentclass[%
 jmp,
 amsmath,amssymb,
 preprint,%
]{revtex4-1}

\usepackage{graphicx}% Include figure files
\usepackage{dcolumn}% Align table columns on decimal point
\usepackage{bm}% bold math
\usepackage{esint}
\usepackage[mathlines]{lineno}% Enable numbering of text and display math

\usepackage[utf8]{inputenc}
\usepackage[T1]{fontenc}
\usepackage{mathptmx}
\usepackage{mathtools}

\newcommand{\abs}[1]{\left\vert #1 \right\vert}
\DeclarePairedDelimiter\norm{\lVert}{\rVert}%

\DeclareMathOperator{\csch}{csch}

\newcommand{\coeffD}{k^2(\frac{\epsilon}{2}k^2\sinh{k^3} + 2\sqrt{\frac{\epsilon}{2}}k\cosh{k^3} + \sinh{k^3})}
\newcommand{\coeffDb}{\frac{\epsilon}{2}k^2\sinh{k^3} + 2\sqrt{\frac{\epsilon}{2}}k\cosh{k^3} + \sinh{k^3}}
\newcommand{\coeffH}{\sqrt{\frac{k}{2}}\Ray\norm{\theta_k}_2}

\renewcommand{\vec}[1]{\boldsymbol{#1}}
\newcommand{\Ra}{\mbox{\it Ra}}
\newcommand{\Ray}{\widetilde{\mbox{\it Ra}}}
\newcommand{\Ek}{\mbox{\it Ek}}

\newcommand{\Nu}{\mbox{\it Nu}}

\def\x{\mbox{\boldmath $x$}}

\usepackage{xcolor}
\definecolor{matlabblue}{RGB}{19,42,139}
\definecolor{matlabred}{RGB}{255,0,0}
\definecolor{matlabgreen}{RGB}{0,126,0}

\usepackage{hyperref}
\hypersetup{colorlinks=true,
            citecolor=blue,
            linkcolor=black,
            breaklinks=true}

\begin{document}

\preprint{AIP/123-QED}

\title[Sample title]{Rigorous bounds on the heat transport of rotating convection with Ekman pumping.}
% Force line breaks with \\

\author{B. Pachev}
%Lines break automatically or can be forced with \\
\author{J. P. Whitehead}%
 \email{whitehead@mathematics.byu.edu.}
\affiliation{Mathematics Department, Brigham Young University%\\This line break forced with \textbackslash\textbackslash
}%

\author{G. Fantuzzi}
\affiliation{Department of Aeronautics, Imperial College London%\\This line break forced% with \\
}%

\author{I. Grooms}
\affiliation{Applied Mathematics Department, University of Colorado Boulder}

\date{\today}% It is always \today, today,
             %  but any date may be explicitly specified

\begin{abstract}
We establish rigorous upper bounds on the time-averaged heat transport for a model of rotating Rayleigh-B\'enard convection between no-slip boundaries at infinite Prandtl number and with Ekman pumping.  The analysis is based on the asymptotically reduced equations derived for rotationally constrained dynamics with no-slip boundaries, and hence includes a lower order correction that accounts for the Ekman layer and corresponding Ekman pumping into the bulk.  Using the auxiliary functional method we find that, to leading order, the temporally averaged heat transport is bounded above as a function of the Rayleigh and Ekman numbers $\Ra$ and $\Ek$ according to $\Nu \leq 0.3704 \Ra^2 \Ek^2$.  Dependent on the relative values of the thermal forcing represented by $\Ra$ and the effects of rotation represented by $\Ek$, this bound is both an improvement on earlier rigorous upper bounds, and provides a partial explanation of recent numerical and experimental results that were consistent yet surprising relative to the previously derived upper bound of $\Nu \lesssim \Ra^3 \Ek^4$.
\end{abstract}

\maketitle

\section{Introduction}
The turbulent motion of a fluid driven by an unstable density gradient is ubiquitous in the natural and engineering sciences \cite{Ah2009,AhGrLo2009}.  The canonical mathematical model of this phenomenon is Rayleigh-B\'enard convection \cite{Ra1916} in which an incompressible fluid is constrained between two horizontal plates typically taken with horizontally periodic boundary conditions.  A natural extension of the original Rayleigh-B\'enard problem is to consider the influence of rigid body rotation about the gravitational axis.  Such rotational effects and their impact on the underlying turbulent evolution are an essential influence in planetary atmospheres, planetary and stellar interiors, stars and terrestrial oceans (see, for example, \cite{Mi2005,gastine2014zonal,aurnou2015rotating,heimpel2016simulation,marshall1999open}).  Investigations into the effects of rotation on the thermally driven turbulent motion are carried out via experiments \cite{rossby1969study,ecke2014heat,cheng2015laboratory,rajaei2017exploring}, simulations \cite{julien2012statistical,stellmach2014approaching,alards2019sharp,toselli2019effects}, and via mathematically motivated asymptotics \cite{julien1998new,julien2016nonlinear,plumley2016effects}, but there remain many unanswered questions.  In particular although there are theoretical predictions of the functional dependence of the time-averaged heat transport on the thermal forcing, rate of rotation, and kinematic properties of the fluid (see \cite{plumley2018scaling} for a recent review), there is no conclusive evidence to determine which of these theories (if any) are correct.

The enhancement of the time-averaged heat transport is measured by the non-dimensional Nusselt number $\Nu$. A fundamental problem that has received increasing attention in recent years is to determine the functional dependence of $\Nu$ on the three other non-dimensional parameters of the system, namely, the Rayleigh, Ekman, and Prandtl numbers (in addition to the geometry of the spatial domain). These are defined, respectively, as
\begin{equation}
    \Ra = \frac{g\alpha (\Delta T)h^3}{\nu \kappa}, \qquad 
    \Ek = \frac{\nu}{2\Omega h^2},\qquad 
    \Pr = \frac{\nu}{\kappa},
\end{equation}
%These dimensionless parameters are formed from the kinematic viscosity $\nu$, thermal diffusivity $\kappa$, gravitational acceleration $g$, thermal expansion coefficient $\alpha$, linear enforced temperature gradient $\Delta T$, vertical extent of the domain $h$, and rigid body rotation rate $\Omega$. 
where $\nu$ is the kinematic viscosity, $\kappa$ is the thermal diffusivity, $g$ is the gravitational acceleration, $\alpha$ is the thermal expansion coefficient, $\Delta T$ is the linear enforced temperature gradient, $h$ is the vertical extent of the domain, and $\Omega$ is the rigid body rotation rate. 

Rigorous upper bounds on the heat transport%, while limited in derivation, 
provide hard limits on the influence of both rotation and thermal forcing on the heat transport (see, for example, \cite{doering1994variational,doco1996}).  Methods to generate such bounds traditionally rely on energy inequalities derived from the Boussinesq equations. Since rigid body rotation does not influence the domain averaged energy, these inequalities and the bounds implied by them do not capture the effects of rotation.  For this reason here we follow an alternative approach and apply upper bound analysis to asymptotically reduced equations, such as those derived in \cite{julien1998new}, that capture the influence of rapid rotation.  In \cite{Grooms_2014} it is shown that $\Nu \leq 20.56 \Ra^3 \Ek^4$ for the asymptotically reduced equations under the constraint of a stress-free boundary at the upper and lower surfaces and with the assumption of infinite $\Pr$ (a valid assumption, for instance, for the Earth's mantle).

In this article, we extend the rigorous bound derived in \cite{Grooms_2014} to the case of no-slip boundary conditions.
Since the model studied in \cite{julien1998new,Grooms_2014} is valid only for stress-free boundaries,
here we consider a different set of asymptotically reduced equations that incorporate the effects of a no-slip boundary condition \cite{julien2016nonlinear,plumley2016effects}. Using the auxiliary functional method \cite{chernyshenko2014polynomial} we show that when the Prandtl number is infinite these equations satisfy $\Nu \leq 0.3704 \Ra^2 \Ek^2$. This bound applies to leading order when $\Ra\Ek \gg 1$, and under the appropriate restrictions on $\Ra$ and $\Ek$ for which the asymptotically reduced model equations are valid.

The rest of this article is organized as follows.  Section \ref{sec:setup} introduces the asymptotically reduced system for rotationally constrained convection with no-slip boundaries at infinite $\Pr$. In section \ref{s:af-method} we apply the auxiliary functional method to formulate a variational problem whose solution yields an upper bound on the mean vertical heat transport. 
%Section \ref{sec:velocity} introduces an explicit formula for the velocity components $w_k(z)$ in terms of the temperature fluctuations $\theta_k(z)$ and derives some preliminary bounds on the elements of this solution that arise from the Ekman pumping non-homogeneous boundary conditions. 
Section \ref{sec:velocity} introduces an explicit formula for the coupling between the vertical velocity and temperature fluctuations around the horizontal mean, and derives preliminary estimates for this coupling that are key to proving our bound on the Nusselt number $\Nu$.
Section \ref{sec:final_bound} combines these estimates with previous analysis from \cite{Grooms_2014} to prove the final rigorous upper bound
\begin{equation}
\label{e:final-bound-Nu}
    %\Nu \leq \frac{10}{27}\Ray^2\epsilon^{-2} = \frac{10}{27}\Ra^2 \Ek^2.
    \Nu \leq \frac{10}{27}\Ra^2\Ek^{2} + \frac{2}{9\sqrt{\gamma}}\left( \frac{\gamma}{2} + 4 \coth{2\gamma} + \frac{\csch^2(2\gamma)}{\gamma}\right)^\frac12 \Ra\,\Ek - \frac{1}{3},
\end{equation}
where $\gamma = \sinh^{-1}(1)$.
Section \ref{sec:conc} discusses this result and its implications for the rotating convection problem. Readers who are interested in the results and an overview of the methodology may restrict their attention to Sections \ref{sec:setup} and \ref{sec:conc}.  Sections \ref{sec:velocity} and \ref{sec:final_bound} contain the technical results leading to~\eqref{e:final-bound-Nu} and can be omitted if only an understanding of the physical implications is desired.

%%%%%%%%%%%%%%%%%%%%%%%%%%%%%%%%%%%
\section{The model}\label{sec:setup}

For notational simplicity, define the scaled Rayleigh and Ekman numbers
\begin{align}
    \widetilde{\Ra} &= \Ra \Ek^{4/3}, &
    \epsilon &= \Ek^{1/3}.
\end{align}
The asymptotically reduced rapidly rotating Rayleigh-B\'enard system with Ekman pumping at infinite Prandtl numbers is described by the non-dimensional equations~\citep{julien2016nonlinear}
\begin{subequations}
\begin{equation}
    \label{eq:model-1}
    \partial_z \psi = \Ray\,\theta + \nabla_h^2 w,
\end{equation}
\begin{equation}
    \label{eq:model-2}
    -\partial_z w = \nabla_h^2\zeta,
\end{equation}
\begin{equation}
    \label{eq:model-3}
    \nabla_h^2 \psi = \zeta,
\end{equation}
\begin{equation}
    \label{eq:model-6}
    \nabla_h\cdot\boldmath{u}_1 + \partial_z w = 0,
\end{equation}
\begin{equation}
    \label{eq:model-4}
    \partial_t\theta + J[\psi,\theta] + w\partial_z\overline{T} + \epsilon\left[\nabla_h \cdot (\boldmath{u}_1\theta) + \partial_z (w\theta)'\right] = \nabla_h^2\theta + \epsilon^2 \partial_z^2\theta,
\end{equation}
\begin{equation}
    \label{eq:model-5}
    \epsilon^{-2}\partial_t\overline{T} + \partial_z\left(\overline{w\theta} \right) = \partial_z^2\overline{T}
\end{equation}
\end{subequations}
Here $\overline{T}$ is the horizontally (and fast-time scale) averaged temperature field, $\theta$ is the fluctuation about this value (hence, it has zero horizontal mean),  $w$ is the vertical velocity, $\psi$ is the stream function of the horizontal velocity (which, in geostrophic balance, coincides with the scalar pressure field), $\boldmath{u}_1$ is an order-$\epsilon$ correction to the horizontal velocity, and $\zeta$ is the vertical component of vorticity.
The notation $\left(\overline{w\theta}\right)$ refers to the horizontal average of the product $w\theta$ of the vertical velocity and temperature fluctuations, while $(w\theta)'$ = $w\theta - \left(\overline{w\theta}\right)$ is the corresponding fluctuation. Note that, at each time instant, the variables $\psi$, $\zeta$, $w$ and $u_1$ can be determined uniquely as a function of the temperature variables $\overline{T}$ and $\theta$ using~\eqref{eq:model-1}--\eqref{eq:model-6}. Thus, $\overline{T}$ and $\theta$ are the only dynamical variables in the problem.

All variables are assumed to have period $L_x$ and $L_y$ in the horizontal $x$- and $y$-directions, respectively. The flow is driven by differential heating in the vertical direction, with $\overline{T} = 1$ at the bottom boundary ($z=0$) and $\overline{T}=0$ at the top boundary ($z=1$). The temperature fluctuations $\theta$ and all horizontal velocity components satisfy Dirichlet boundary conditions at both vertical boundaries, while the vertical velocity field satisfies the Ekman pumping conditions
\begin{align}
    \nabla_h^2w &= \sqrt{\frac{\epsilon}{2}}\partial_z w \quad\text{at}~z=1,
    &
    \nabla_h^2w &= -\sqrt{\frac{\epsilon}{2}}\partial_z w \quad\text{at}~z=0.
\end{align}

We are interested in deriving rigorous \textit{a priori} bounds for the Nusselt number $\Nu$, which is defined as
\begin{equation}\label{eq:Nu_defn}
    \Nu := \limsup_{t_0\rightarrow\infty}\frac{1}{t_0}\int_0^{t_0} \mathcal{N}\!\left\{\overline{T}(\cdot,t), \theta(\cdot,t)\right\} dt,
\end{equation}
where
\begin{equation}
\mathcal{N}\!\left\{\overline{T},\theta\right\} :=
    \fint_{\Omega} \left[
    \vert \overline{T}_z \vert^2 
    + \vert \nabla_h\theta \vert^2 + \epsilon^2 \vert \theta_z \vert^2
    \right] d\vec{x}.
\end{equation}
Here and elsewhere the notation $\fint_\Omega(\cdot) d\vec{x}$ denotes averages over the domain.

%%%%%%%%%%%%%%%%%%%%%%%%%%%%%%%%%%%%%%%%%
\section{Bounding the Nusselt number}
\label{s:af-method}

To bound the Nusselt number $\Nu$ we follow a general approach~\cite{chernyshenko2014polynomial,fantuzzi2018bounds,tobasco2018optimal,goluskin2019bounds} and look for a so-called \textit{auxiliary functional} of the dynamical variables, $\mathcal{V}\!\left\{\overline{T},\theta\right\}$, that is differentiable along solutions of~\eqref{eq:model-1}--\eqref{eq:model-5}, uniformly bounded in time, and satisfies
\begin{equation}
    \label{eq:bounding-condition-general}
    U - \mathcal{N}\!\left\{\overline{T}(\cdot,t), \theta(\cdot,t)\right\} - \frac{d}{dt}\mathcal{V}\!\left\{\overline{T}(\cdot,t), \theta(\cdot,t)\right\} \geq 0
\end{equation}
for some constant $U$ and all solutions of~\eqref{eq:model-1}--\eqref{eq:model-5}. Averaging this inequality over time yields the upper bound $\Nu \leq U$ because, by virtue of the boundedness of $\mathcal{V}$,
\begin{equation}
    \limsup_{t_0\rightarrow\infty}\frac{1}{t_0}\int_0^{t_0} \frac{d}{dt}\mathcal{V}\!\left\{\overline{T}(\cdot,t), \theta(\cdot,t)\right\} = 0.
\end{equation}

As shown in~\cite{chernyshenko2017relationship}, this approach generalizes the traditional ``background method'' by Doering \& Constantin (see, e.g.,~\cite{doering1994variational}) as well as a more recent approach by Seis \& Otto~\cite{Otto2011rayleigh,Seis2015scaling}, both of which correspond to $\mathcal{V}$ being quadratic. Optimization of more general $\mathcal{V}$ may produce bounds that are sharp or nearly so~\cite{tobasco2018optimal,goluskin2019bounds}, but such formulations are more difficult to deal with rigorously. For this reason, we will consider the following simple choice for $\mathcal{V}$, which corresponds to applying the background method to the temperature fluctuation $\theta$:
\begin{equation}\label{eq:AF}
\mathcal{V}\!\left\{\overline{T}, \theta\right\} = \int_\Omega\left[\frac{b}{2\epsilon^2}|\Theta|^2 + \frac{b}{2}|\theta|^2 - \frac{1}{\epsilon^2}\phi(z) \Theta \right]d\x,
\end{equation}
where
\begin{equation}
    \Theta(z) := \overline{T}(z) - 1 + z
\end{equation}
is the temperature fluctuation from the conductive profile $\overline{T}_c(z) = 1-z$. Working with $\Theta$ is convenient because it satisfies homogeneous boundary conditions,
\begin{equation}
    \Theta(0)=0=\Theta(1).
\end{equation}
In order to derive the best bound on $\Nu$ that this approach has to offer, the scalar $b$ and the function $\phi(z)$ are to be chosen such that~\eqref{eq:bounding-condition-general} holds with the smallest possible $U$.

The time derivative of $\mathcal{V}\!\left\{\overline{T}, \theta\right\}$ can be calculated using~\eqref{eq:model-1}--\eqref{eq:model-5} and integrating by parts using the boundary conditions, the identity $(w\theta)'=w\theta - \overline{w\theta}$, and the definition of $\theta$ as the perturbation from the horizontally-averaged temperature, so $\overline{\theta}=0$. We find
\begin{multline}\label{e:dVdt-gio}
\frac{d}{dt} \mathcal{V}\!\left\{\overline{T}, \theta\right\} = \int_\Omega\left[-b (|\Theta_z|^2 + |\nabla_h\theta|^2 + \epsilon^2|\theta_z|^2) + (b - \phi')w\theta + \phi' \Theta_z \right]d\x,
\\
- \phi(1)\Theta_z(1) + \phi(0)\Theta_z(0),
\end{multline}
where primes denote differentiation with respect to $z$. Then,~\eqref{eq:bounding-condition-general} requires that the functional
\begin{multline}
    \mathcal{S} :=
    U - 1
    + (b-1)\int_\Omega 
         \left(|\Theta_z|^2 + |\nabla_h\theta|^2 + \epsilon^2|\theta_z|^2 \right) d\x
         \\
        + \int_\Omega \left[(\phi' - b)w\theta - \phi' \Theta_z \right] d\x
    + \phi(1)\Theta'(1) - \phi(0)\Theta'(0)
\end{multline}
be non-negative along all solutions to~\eqref{eq:model-1}--\eqref{eq:model-5}. 

At this stage, we make a key simplification: we drop the time-dependent equations~\eqref{eq:model-4} and~\eqref{eq:model-5}, and enforce the sufficient condition that $\mathcal{S}$ be non-negative for all trial fields $\Theta$ and $\theta$ satisfying the boundary conditions, with $w$ determined by~\eqref{eq:model-1}--\eqref{eq:model-6}.
A standard Fourier decomposition in the two horizontal directions (see, e.g., \cite{Grooms_2014}) reveals that this sufficient condition is equivalent to two inequalities. The first is
\begin{equation}
\label{eq:S0-inequality}
\mathcal{S}_0 := U-1 + \int_0^1\left[(b-1)\left(|\Theta'|^2 + \epsilon^2|\theta_0'|^2\right) - \phi'\Theta'\right] dz
+ \phi(1)\Theta'(1) - \phi(0)\Theta'(0)
\geq 0,
\end{equation}
which should be satisfied for all $z$-dependent fields $\Theta$ and $\theta_0$ satisfying homogeneous Dirichlet boundary conditions. The second is

\begin{equation}
\label{eq:Sk-inequality}
\mathcal{S}_k := \int_0^1\left[(b-1)\left( k^2|\theta_k|^2 + \epsilon^2|\theta_k'|^2 \right) - (b-\phi')w_k\theta_k \right]dz \geq 0,
\end{equation}
imposed for all wavenumbers $k$ and all $z$-dependent fields $\theta_k$, $w_k$ related by

\begin{equation}
%    \tag{Temperature Velocity Slaving}
    k^6w_k - w_k''= k^4\Ray\theta_k
    \label{eqn:slaving}
\end{equation}
and subject to the boundary conditions 
\begin{align}
\label{e:bcs-fourier}
    \theta_k(0) &= 0, &
    \theta_k(1) &= 0, &
    k^2w_k(0) &= \sqrt{\frac{\epsilon}{2}}w'_{k}(0), &
    k^2w_k(1) &= -\sqrt{\frac{\epsilon}{2}}w'_{k}(1).
\end{align}

Boundedness of $\mathcal{S}_0$ from below requires that the sign-indefinite boundary terms in~\eqref{eq:S0-inequality} vanish, so we set $\phi(0)=\phi(1)=0$. Further, minimizing $\mathcal{S}_0$ over $\Theta$ and $\theta_0$ using the Calculus of Variations shows that
\begin{equation}
    \label{eqn:U_bound}
    U = 1 + \frac{\norm{\phi'}^2_2}{4(b-1)}
\end{equation}
is the smallest value such that $\mathcal{S}_0\geq 0$.
Thus, the problem reduces to choosing the function $\phi(z)$ and the scalar $b$ that satisfy~\eqref{eq:Sk-inequality} and minimize the right-hand side of~\eqref{eqn:U_bound}.

\begin{figure*}
    \centering
    \includegraphics[scale=1]{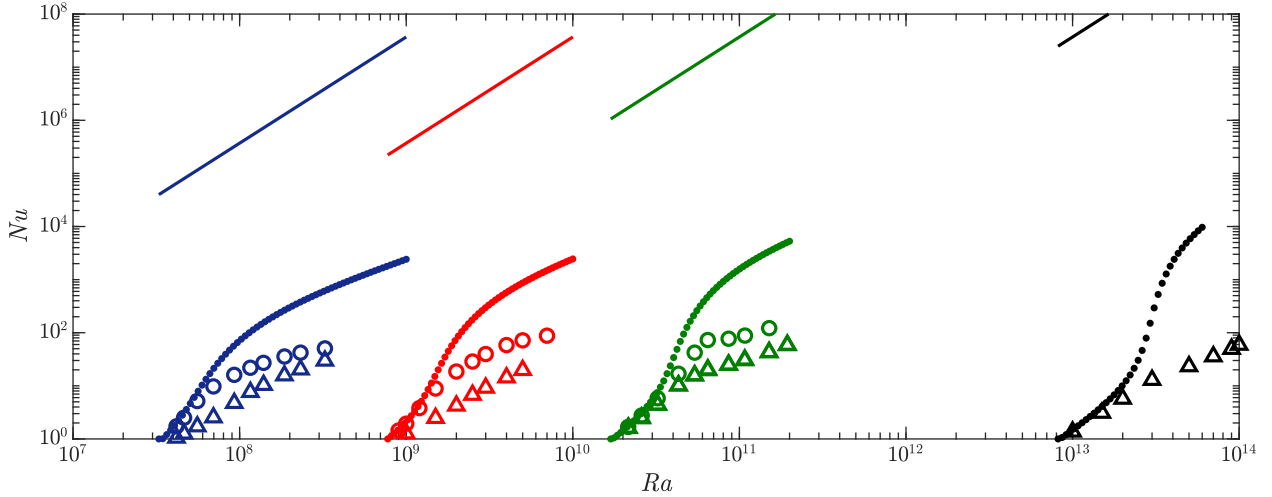}
    \caption{Upper bounds on $\Nu$ for 
    $\Ek = \epsilon^3 = 10^{-5}$
    ({\color{matlabblue}\footnotesize$\bullet$}~optimized, 
     {\color{matlabblue}\bf ---}~analytical),
    $10^{-6}$
    ({\color{matlabred}\footnotesize$\bullet$}~optimized,
     {\color{matlabred}\bf ---}~analytical),
    $10^{-7}$
    ({\color{matlabgreen}\footnotesize$\bullet$}~optimized,
     {\color{matlabgreen}\bf ---}~analytical)
    and
    $10^{-9}$
    ({\color{black}\footnotesize$\bullet$}~optimized,
    {\color{black}\bf ---}~analytical),
    compared to DNS data for $\Pr=1$
    ({\footnotesize{\color{matlabblue}$\boldsymbol{\triangle}$}} for $\Ek = 10^{-5}$,
    {\footnotesize{\color{matlabred}$\boldsymbol{\triangle}$}} for $\Ek = 10^{-6}$,
    {\footnotesize{\color{matlabgreen}$\boldsymbol{\triangle}$}} for $\Ek = 10^{-7}$,
    {\footnotesize{\color{black}$\boldsymbol{\triangle}$}} for $\Ek = 10^{-9}$)
    and $\Pr = 7$
    ({\large{\color{matlabblue}$\boldsymbol{\circ}$}} for $\Ek = 10^{-5}$,
    {\large{\color{matlabred}$\boldsymbol{\circ}$}} for $\Ek = 10^{-6}$,
    {\large{\color{matlabgreen}$\boldsymbol{\circ}$}} for $\Ek = 10^{-7}$).}
    \label{fig:optimized-bounds}
\end{figure*}

\begin{figure*}
    \centering
    \includegraphics[scale=1]{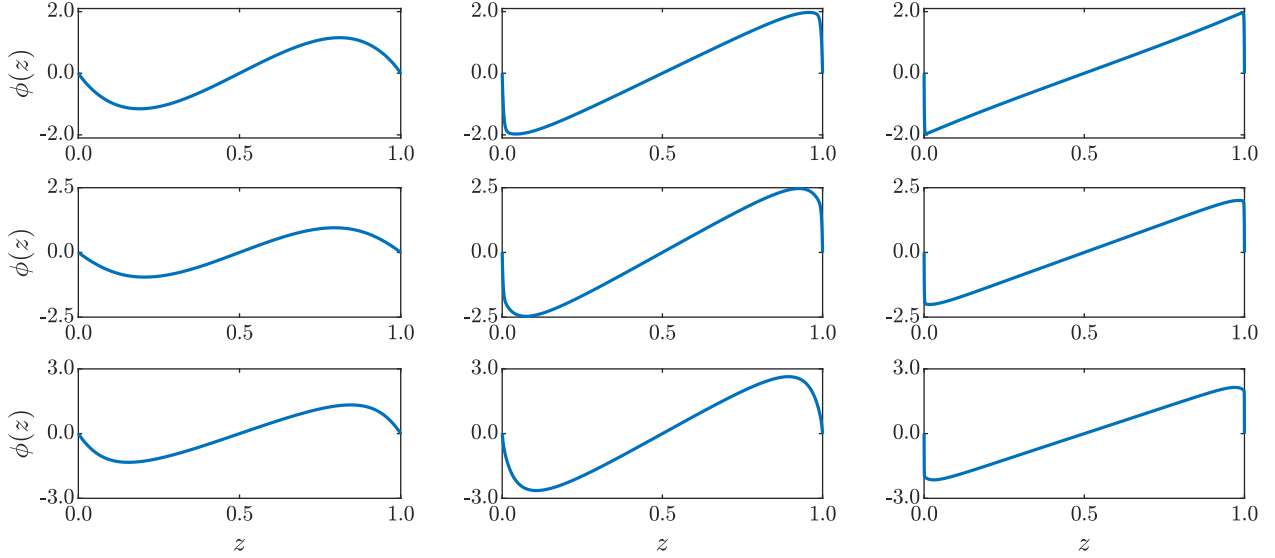}
    \caption{Optimal profiles $\phi(z)$.
    \textit{Top:} $\Ek=10^{-5}$ and
    (a)~$\Ra \approx 4.0684 \times 10^7$,
    (b)~$\Ra \approx 9.1825 \times 10^7$,
    (c)~$\Ra \approx 2.7186 \times 10^8$.
    \textit{Middle:} $\Ek=10^{-6}$ and
    (d)~$\Ra \approx 8.9022 \times 10^8$,
    (e)~$\Ra \approx 1.8738 \times 10^9$,
    (f)~$\Ra \approx 3.5938 \times 10^9$.
    \textit{Bottom:} $\Ek=10^{-7}$ and
    (g)~$\Ra \approx 2.5505 \times 10^{10}$,
    (h)~$\Ra \approx 3.5136 \times 10^{10}$,
    (i)~$\Ra \approx 8.3828 \times 10^{10}$.}
    \label{f:optimal-profiles}
\end{figure*}

The optimal $\phi$ and $b$ can be estimated numerically after approximating this infinite-dimensional optimization problem with a finite-dimensional semidefinite program---a well-known type of convex optimization problem. This can be done in various ways; in this work, we apply the Legendre series expansion method described in~\cite{fantuzzi2016pre,fantuzzi2017optimization}, which is implemented in the MATLAB toolbox {\sc quinopt}~\cite{fantuzzi2017quinopt}. To avoid large powers of the wavenumber $k$ when $\theta_k$ is expressed in terms of $w_k$ using~\eqref{eqn:slaving}, we modified {\sc quinopt} to explicitly solve for (the expansion coefficients of) $w_k$ in terms of $\theta_k$; our code is available at \url{https://github.com/aeroimperial-optimization/rotating-convection-EP}.
%, or by solving the Euler-Lagrange optimality conditions using the time-marching strategy by Wen \textit{et al.}~\cite{Wen2013computational,Wen2015timestepping}. 
Figure~\ref{fig:optimized-bounds} compares bounds optimized in this way for $\Ek = \epsilon^3 = 10^{-5}$, $10^{-6}$ and $10^{-7}$ to available DNS data for finite $\Pr=1$ and $7$~\cite{plumley2016effects,plumley2017sensitivity,cheng2015laboratory}, as well as the rigorous bound~\eqref{e:final-bound-Nu}. Optimal profiles $\phi(z)$ corresponding to the optimal bounds of figure~\ref{fig:optimized-bounds}, which are shown in figure~\ref{f:optimal-profiles} for a selection of Ekman numbers $\Ek$ and Rayleigh numbers $\Ra$.

To prove the analytical bound~\eqref{e:final-bound-Nu}, we model the optimal $\phi(z)$ using the piecewise linear profile
\begin{equation}
    \phi(z) = \begin{cases}-\frac{c}{2}(\frac{1-2\delta}{\delta})z & 0 \le z \le \delta\\
    c(z-\frac{1}{2})  & \delta \le z \le 1-\delta\\
    \frac{c}{2}(\frac{1-2\delta}{\delta})(1-z) & 1 -\delta \le z \le 1,
    \end{cases}
\end{equation}
where $c$ and $\delta$ are positive scalars. With this choice, the upper bound on the Nusselt number becomes
\begin{equation}
\label{e:analytical-bound-formula}
    \Nu \leq 1 + \frac{\norm{\phi'}^2_2}{4(b-1)} = 1 + \frac{b^2}{4(b-1)}\left(\frac{1}{2\delta}-1\right),
\end{equation}
while, after some algebra,~\eqref{eq:Sk-inequality} requires that
\begin{multline}
\label{e:analytical-Sk}
    (b-1)\varepsilon^2 \|\theta_k'\|_2^2 + (b-1)k^2 \|\theta_k\|_2^2 \\
    - (b-c)\int^{1}_{0}\theta_k w_k dz 
    - \frac{c}{2\delta}\Bigg[\int^{\delta}_{0}\theta_k w_k dz + \int^{1}_{1-\delta}\theta_k w_k dz\Bigg]
    \geq 0.
\end{multline}
for all wavenumbers $k > 0$.
In the rest of this work we will use~\eqref{e:analytical-bound-formula} prove~\eqref{e:final-bound-Nu} by choosing $b$, $c$ and $\delta$ such that~\eqref{e:analytical-Sk} holds.

\section{Understanding the velocity field}\label{sec:velocity}
To establish~\eqref{e:analytical-Sk}
we need to bound the indefinite term in the integral which involves the cross term $\theta_kw_k$. There are two standard approaches to doing this.  One is to use \eqref{eqn:slaving} to rewrite $\theta_k$ in terms of $w_k$. The other, which we pursue here, is to note that \eqref{eqn:slaving} is linear and can be explicitly solved to obtain $w_k$ in terms of $\theta_k$. The general Green's function solution for \eqref{eqn:slaving} is
\begin{equation}
     w_k(z) = c_1\sinh(k^3z) + c_2\cosh{k^3z} + h_k(z)
\end{equation}
where $c_1$ and $c_2$ are constants determined by the boundary conditions, while $h_k(z)$ is the solution of \eqref{eqn:slaving} when the velocity $w_k$ is subject to the Dirichlet boundary conditions $w_k(0)=0=w_k(1)$ instead of the Ekman pumping conditions of Robin type in~\eqref{e:bcs-fourier}. Precisely,
\begin{equation}
\label{e:hk}
    h_k(z) = \frac{\Ray k}{\sinh{k^3}}\int^{1}_{0} g(z,s)\theta_k(s)ds
\end{equation}
  where 
  \begin{equation}
      g(z,s) = \begin{cases} 
      \sinh{k^3z}\sinh{k^3(1-s)} & z \leq s \\
      \sinh{k^3s}\sinh{k^3(1-z)} & s \leq z. 
   \end{cases}
\end{equation}

As in \cite{Grooms_2014}, the function $h_k(z)$ can be bounded for all $k$. It remains to identify and then bound the coefficients $c_1$ and $c_2$. In particular, we need to show that $c_1+c_2$ decays at a rate of $e^{-k^3}$ for large $k$ to ensure that the solution is controlled at the smallest scales. This is necessary to balance the $e^{k^3z}$ terms from $\sinh{}$ and $\cosh{}$, which become apparent when $w_k$ is rewritten as 
 \begin{equation}
    \label{e:wk-analytical-solution}
     w_k(z) = \frac{c_1+c_2}{2}e^{k^3z} + \frac{c_2-c_1}{2}e^{-k^3z} + h_k(z) = Ae^{k^3z} + Be^{-k^3z} + h_k(z),
 \end{equation}
 where we have introduced the constants $A$ and $B$ (dependent on $c_1$ and $c_2$) for convenience.
 
 Recalling that $h_k$ satisfies Dirichlet boundary conditions, we have
 \begin{align}
    w_k(0) &= c_2, \\ 
    w_k'(0) &= k^3c_1 + h_k'(0), \\
    w_k(1) &= c_1\sinh(k^3) + c_2\cosh(k^3), \\
    w_k'(1) &= k^3(c_1\cosh{k^3} + c_2\sinh{k^3}) + h_k'(1).
\end{align}
Using the boundary conditions for $w_k$ in \eqref{e:bcs-fourier} and regrouping terms, we obtain
\begin{align}
    -k\sqrt{\frac{\epsilon}{2}}c_1 + c_2&  = \frac{1}{k^2}\sqrt{\frac{\epsilon}{2}}h_k'(0), \\
    \left(\sinh{k^3} + k\sqrt{\frac{\epsilon}{2}}\cosh{k^3}\right)c_1 + \left(\cosh{k^3} + k\sqrt{\frac{\epsilon}{2}}\sinh{k^3}\right)c_2 & = \frac{-1}{k^2}\sqrt{\frac{\epsilon}{2}}h_k'(1).
\end{align}
Solving for $c_1$ and $c_2$ yields
\begin{align}
        c_1 &= -\frac{(\frac{\epsilon}{2}k\sinh{k^3} + \sqrt{\frac{\epsilon}{2}}\cosh(k^3))h_k'(0) + \sqrt{\frac{\epsilon}{2}}h_k'(1)}{\coeffD}, \\
        c_2 &= \frac{(\frac{\epsilon}{2}k\cosh{k^3} + \sqrt{\frac{\epsilon}{2}}\sinh(k^3))h_k'(0) - \frac{\epsilon}{2}k h_k'(1)}{\coeffD}.
\end{align}

The rest of this Section is dedicated to providing key estimates on the velocity $w_k(z)$ both for large and small values of $k$.  As \cite{Grooms_2014} already established estimates on $h_k(z)$, we focus on bounding the exponential terms in \eqref{e:wk-analytical-solution}.  These estimates yield sufficient control over the velocity, such that the integrals depending on $w_k(z)\theta_k(z)$ in~\eqref{e:analytical-Sk} are dominated by the other positive definite terms.

\subsection{Analysis for small scales: $k > 1$}
To estimate the effects of the vertical velocity components $w_k$ for wavenumbers $k > K$, we return to the exponential definition of $w_k$.  The critical wavenumber $K$ becomes an additional parameter to optimize the final bound over, but different values of $K$ will only slightly improve the prefactor in the final bound, and so we content ourselves with $K=1$ to simplify the computations that follow.
The coefficient $A$ of $e^{k^3z}$ in \eqref{e:wk-analytical-solution} is given by
\begin{equation}
    A = \frac{c_1+c_2}{2} = \frac{\left(\frac{\epsilon}{2}k(\cosh{k^3}-\sinh{k^3}) + \sqrt{\frac{\epsilon}{2}}(\sinh{k^3}-\cosh{k^3})\right)h_k'(0) - \left(k\frac{\epsilon}{2}+\sqrt{\frac{\epsilon}{2}}\right)h_k'(1)}{2\coeffD}.
\end{equation}
Upon simplifying terms, we obtain
\begin{equation}
    \label{e:A-coeff}
    A = \frac{\left(\frac{\epsilon}{2}k-\sqrt{\frac{\epsilon}{2}}\right)e^{-k^3}h_k'(0) - \left(k\frac{\epsilon}{2}+\sqrt{\frac{\epsilon}{2}}\right)h_k'(1)}{2\coeffD}.
\end{equation}
For large $k$, this is $O\left(\frac{e^{-k^3}}{k^3}\right)h_k'(1)$. %which is a sufficiently strong decay in $k$. 
Similarly,
\begin{equation}
    \label{e:B-coeff}
    B = \frac{c_2-c_1}{2} = \frac{(k\frac{\epsilon}{2}+\sqrt{\frac{\epsilon}{2}})e^{k^3}h_k'(0) + (\sqrt{\frac{\epsilon}{2}}-k\frac{\epsilon}{2})h_k'(1)}{2\coeffD},
\end{equation}
which is of order $\frac{h_k'(0)}{k^3}$ for $k \gg 1$. Thus, provided that $h_k'(0)$ and $h_k'(1)$ can be controlled for large $k$, the terms $\frac{c_1+c_2}{2}e^{k^3z} + \frac{c_2-c_1}{2}e^{-k^3z}$ in $w_k$ will be bounded for all $k > 1$.% (note that the small $k$ limit is not problematic in this formulation of $w_k$).

Let us now estimate $h_k'(0)$ and $h_k'(1)$ for $k > 1$. Upon breaking up the integral in~\eqref{e:hk} at $z$ and differentiating with Leibniz's rule, we find that the derivative of $h_k$ is
\begin{equation}
    h_k'(z) = 
    \frac{\Ray\, k^4}{\sinh{k^3}}\left[
    \int_{z}^{1}\cosh{k^3z}\sinh{k^3(1-s)}\theta_k(s) ds 
    - \int_0^z\sinh{k^3s}\cosh{k^3(1-z)}\theta_k(s) ds
    \right].
\end{equation}
By the Cauchy-Schwarz inequality, we obtain
\begin{gather}
\label{e:bound-hk-prime-at-0}
    \abs{h_k'(0)} 
    \le \Ray\, k^4 \int_0^1 \frac{\abs{ \sinh{k^3(1-s)} \theta_k(s) }}{\sinh{k^3}}  ds
    \le \frac{\Ray \,k^4\norm{\theta_k}_2}{2\sinh{k^3}}\sqrt{\frac{\sinh{2k^3}}{k^3}-2},
    \\
\label{e:bound-hk-prime-at-1}
    \abs{h_k'(1)} 
    \le \Ray\, k^4
    \int_0^1 \frac{\abs{ \sinh{k^3s} \theta_k(s) }}{\sinh{k^3}} ds
    \le \frac{\Ray\, k^4\norm{\theta_k}_2 }{ 2\sinh{k^3} } \sqrt{ \frac{\sinh{2k^3}}{k^3}-2. }   
\end{gather}
%\end{equation}
We can further estimate
\begin{equation}
\label{e:bound-hk-prime-simplification}
\Ray k^{2.5}\norm{\theta_k}_2\sqrt{\frac{\sinh{2k^3}-2k^3}{4\sinh^2{k^3}}} \le \Ray k^{2.5}\norm{\theta_k}_2\frac{\sqrt{2}}{2},
\end{equation}
and this estimate is tight since
\begin{equation}
\lim_{k\to\infty}\sqrt{\frac{\sinh{2k^3}-2k^3}{4\sinh^2{k^3}}} = \frac{\sqrt{2}}{2}.
\end{equation}
The last inequality can be justified by noting that the function $\sinh^2{x}$ is convex for positive $x$, so
\begin{equation}
    \int_0^x\sinh^2{s}ds \le \frac{1}{2}(\sinh^2(x)-\sinh^2{0}) = \frac{1}{2}\sinh^2{x}.
\end{equation}
Thus 
\begin{equation}
    \sinh{2k^3}-2k^3 = 4 \int_0^{k^3}\sinh^2{x}dx \le 2\sinh^2{k^3},
\end{equation}
which implies the desired inequality.

Inserting the bounds \eqref{e:bound-hk-prime-at-0}--\eqref{e:bound-hk-prime-simplification} into \eqref{e:A-coeff} and \eqref{e:B-coeff}, we can estimate the coefficients $A$ and $B$ as
\begin{align}
\label{e:bound-A}
        \abs{A} \le \sqrt{2}k^{\frac{1}{2}}\Ray\norm{\theta_k}_2\frac{\abs{\frac{\epsilon}{2}k-\sqrt{\frac{\epsilon}{2}}}e^{-k^3} +\abs{k\frac{\epsilon}{2}+\sqrt{\frac{\epsilon}{2}}}}{4(\coeffDb)}, \\
\label{e:bound-B}
        \abs{B} \le \sqrt{2}k^{\frac{1}{2}}\Ray\norm{\theta_k}_2\frac{\abs{k\frac{\epsilon}{2}+\sqrt{\frac{\epsilon}{2}}}e^{k^3} + \abs{\sqrt{\frac{\epsilon}{2}}-k\frac{\epsilon}{2}}}{4(\coeffDb).}
\end{align}
These bounds can be significantly simplified as follows. Note that
\begin{align}
%\begin{split}
    \frac{\abs{k\frac{\epsilon}{2}+\sqrt{\frac{\epsilon}{2}}}e^{k^3} + \abs{\sqrt{\frac{\epsilon}{2}}-k\frac{\epsilon}{2}}}{2(\coeffDb)} 
    \hspace{-2.25in}&
    \nonumber\\
    &= \frac{1}{k}\Bigg(1 + k\frac{\abs{k\frac{\epsilon}{2}+\sqrt{\frac{\epsilon}{2}}}e^{k^3} + \abs{\sqrt{\frac{\epsilon}{2}}-k\frac{\epsilon}{2}}}{2(\coeffDb)} - 1\Bigg)
    %\end{split}&&
    \nonumber\\
    &= \frac{1}{k}\Bigg(1 + \frac{k\abs{\sqrt{\frac{\epsilon}{2}}-k\frac{\epsilon}{2}} + (\frac{\epsilon}{2}k^2-
    \sqrt{\frac{\epsilon}{2}}k)e^{-k^3} 
    - 2\sqrt{\frac{\epsilon}{2}}k\cosh{k^3} 
    - 2\sinh{k^3}}{2(\coeffDb)}\Bigg)
    \nonumber\\
    &\le \frac{1}{k}\Bigg(1 + \frac{k\sqrt{\frac{\epsilon}{2}}\left[\abs{1-k\sqrt{\frac{\epsilon}{2}}}(1+e^{-k^3}) - 2\cosh{k^3}\right]}{2(\coeffDb)} \Bigg).\label{eq:fraction_bound}
\end{align}
Since $k > 1 > 0$, we can further estimate part of the numerator of the last expression as
\begin{equation}
    \abs{1-k\sqrt{\frac{\epsilon}{2}}}(1+e^{-k^3}) \le 2\abs{1-k\sqrt{\frac{\epsilon}{2}}} \le 2 \max\Bigg(1,k\sqrt{\frac{\epsilon}{2}}\Bigg).
\end{equation}
Assuming that $\epsilon < 2$ (in fact, $\epsilon \ll 1$ for the flow regime of interest), the previous estimate is bounded above by $2\max(1,k) \le 2\cosh{k^3}$. Thus the fraction in \eqref{eq:fraction_bound} is less than zero and we see that, uniform in $k$,
\begin{equation}
    \frac{\abs{k\frac{\epsilon}{2}+\sqrt{\frac{\epsilon}{2}}}e^{k^3} + \abs{\sqrt{\frac{\epsilon}{2}}-k\frac{\epsilon}{2}}}{2(\coeffDb)} \le \frac{1}{k}.
\end{equation}%
Substituting this estimate back into the bounds for $\abs{A}$ and $\abs{B}$ from \eqref{e:bound-A} and \eqref{e:bound-B} we find
\begin{align}
    \label{e:bound-A-final}
    \abs{A} \le \frac{\sqrt{2}}{2}e^{-k^3}k^{\frac{-1}{2}}\Ray\norm{\theta_k}_2,\\
    \abs{B} \le \frac{\sqrt{2}}{2}k^{\frac{-1}{2}}\Ray\norm{\theta_k}_2.
    \label{e:bound-B-final}
\end{align}

\subsection{Analysis for large scales: $k \leq 1$}
While the exponential formulation is useful for large $k$, the original formulation in terms of $\sinh$ and $\cosh$ is easier to work with when $k$ approaches zero. Here we establish useful bounds on $\abs{c_1}$ and $\abs{c_2}$ which are valid for $k \leq 1$.
Using estimates \eqref{e:bound-hk-prime-at-0} and \eqref{e:bound-hk-prime-at-1} for the value of $h'_k$ at the boundaries, we find
\begin{align}
    \abs{c_1} 
    &\le \coeffH\frac{\frac{\epsilon}{2}k\sinh{k^3} + \sqrt{\frac{\epsilon}{2}}\cosh{k^3} + \sqrt{\frac{\epsilon}{2}}}{\coeffDb}\nonumber \\
    &= \coeffH\Bigg[\frac{1}{2k} + \frac{\sqrt{\frac{\epsilon}{2}} + \frac{\epsilon}{4}k\sinh{k^3} - \frac{1}{2k}\sinh{k^3}}{\coeffDb}\Bigg]\nonumber\\
    &\le \coeffH\Bigg[\frac{1}{2k} + \frac{\sqrt{\frac{\epsilon}{2}}}{2\sqrt{\frac{\epsilon}{2}}k\cosh{k^3}}+\frac{\sinh{k^3}(\frac{\epsilon}{4}k - \frac{1}{2k})}{\coeffDb} \Bigg]\nonumber\\
    &\le \coeffH\Bigg[\frac{1}{k}+\frac{\sinh{k^3}(\frac{\epsilon}{4}k - \frac{1}{2k})}{\coeffDb}\Bigg].
\end{align}

For $k \le 1 \leq \sqrt{\frac{2}{\epsilon}}$ (recall that we are interested in $\varepsilon \ll 1$) we have $\frac{\epsilon}{4}k < \frac{1}{2k}$, so this bound simplifies to
\begin{equation}
 \abs{c_1} \le \coeffH\Bigg(\frac{1}{k}\Bigg) = \sqrt{\frac{1}{2k}}\Ray\norm{\theta_k}_2. 
 \label{eqn:c1_bound} 
\end{equation}
Following a similar analysis for $c_2$ yields
\begin{equation}
\abs{c_2} \le \coeffH\Bigg(\sqrt{\frac{\epsilon}{2}}\Bigg) = \frac{1}{2}\sqrt{\epsilon k}\Ray\norm{\theta_k}_2.
    \label{eqn:c2_bound}
\end{equation}

%%%%%%%%%%%%%%%%%%%%%%%%%%%%%%%%%%%%%%%%%
%\section{One Bound to bound them all}
\section{Proof of the main result}\label{sec:final_bound}
We now return to the original problem of proving~\eqref{e:final-bound-Nu} by showing that (\ref{e:analytical-Sk}) holds for all wavenumbers $k > 0$.
This requires bounding the sign-indefinite terms 
%$f(k) = \int^{1}_{0}(b-\phi'(z))\theta_k(z)w_k(z)$  Plugging in our choice for $\phi$, we obtain
\begin{equation}
%\begin{split}
%    f(k) = 
%    (b-c)\int^{1-\delta}_{\delta}\theta_kw_kdz + \left[b+c\left(\frac{1-2\delta}{2\delta}\right)\right]\Bigg[\int^{\delta}_{0}\theta_kw_kdz + \int^{1}_{1-\delta}\theta_kw_kdz\Bigg]\\
%    = 
    (b-c)\int^{1}_{0}\theta_kw_kdz + \frac{c}{2\delta}\Bigg[\int^{\delta}_{0}\theta_kw_kdz + \int^{1}_{1-\delta}\theta_kw_kdz\Bigg]
%    \label{eqn:f}
\label{e:indefinite-terms}
\end{equation}
in terms of the $L_2$ norm of $\theta_k$ and $\theta_k'$.
Although taking $c$ slightly less than $b$ leads to a slight improvement of lower-order terms in the final bound, the added complexity is not worth the minor improvement. Hence, we let $c=b$ for the remainder of this analysis to eliminate the first term in~\eqref{e:indefinite-terms}. Moreover, for added simplicity we let $\mathcal{B} = [0,\delta]\cup [1-\delta,1]$ be the union of the upper and lower boundary layers. With this notation in hand, the second and third terms in \eqref{e:indefinite-terms} can be broken into two components: 
\begin{equation}
    \frac{b}{2\delta}\int_{\mathcal{B}} \theta_kw_kdz = \frac{b}{2\delta}\int_{\mathcal{B}}\theta_kh_kdz + \frac{b}{2\delta}\int_{\mathcal{B}}\theta_k(Ae^{k^3z}+Be^{-k^3z})dz.
\end{equation}
%In \cite{Grooms_2014} it was shown that the first component can be uniformly dominated by $k^2\norm{\theta_k}_2^2$ for $\delta\sim\Ray^{-3}$. Hence we for now concentrate on the second term. With the insights gained there, we are able to show the contribution from the first term (homogeneous term) is of lower order and has little effect on the final bound.  As before when providing estimates on the velocity field $w_k(z)$ itself, we concentrate on the large $k$ and small $k$ limits separately and then combine the relevant estimates for a uniform bound.
In subsections \ref{s:bl-estimates-large-k}--\ref{s:matching-bounds}, we estimate the contribution of the exponential terms to the boundary integral of $\theta_k w_k$. The boundary integral of $\theta_k h_k$, instead, is estimated in subsection \ref{s:bl-estimates-green}. These estimates are then combined in subsection \ref{s:final-bound} to prove~\eqref{e:final-bound-Nu}.

%%%%%%%%%%%%%%%%%%%%%%%%%%%%%%%%%
\subsection{Contribution of exponential terms for $k > 1$}\label{s:bl-estimates-large-k}
Since $\theta_k(0) = 0 = \theta_k(1)$ we can write 
\begin{equation}
    \theta_k = \int_0^z \theta_k'dz = -\int_z^1\theta_k'dz,
\end{equation}
and by the Cauchy-Schwarz inequality obtain the following piece-wise bound:
\begin{equation}
    \abs{\theta_k} \le \begin{cases} \sqrt{z}\norm{\theta_k'}_2 & 0\le z\le\frac{1}{2}\\
    \sqrt{1-z}\norm{\theta_k'}_2 & \frac{1}{2} \le z \le 1
    \end{cases}.
\end{equation}
Hence, using the fact that $\delta \le \frac{1}{2}$, 
\begin{equation}
    \int_B\theta_k^2dz \le \norm{\theta_k'}_2^2\Bigg[\int_0^{\delta}zdz + \int_{1-\delta}^{1}(1-z)dz\Bigg] = \delta^2\norm{\theta_k'}_2^2.
\end{equation}
In particular, taking $\delta = \frac{1}{2}$, we have $\norm{\theta_k}_2 \le \frac{1}{2}\norm{\theta_k'}_2$.

Further, combining the Cauchy-Schwarz inequality with the estimates for $\abs{A}$ and $\abs{B}$ from \eqref{e:bound-A-final} and \eqref{e:bound-B-final} yields
\begin{align}\nonumber
    \frac{b}{2\delta}\int_{\mathcal{B}}\abs{\theta_k(Ae^{k^3z}+Be^{-k^3z})dz} 
    &\le \frac{b}{2\delta}\int_{\mathcal{B}}\abs{\theta_kAe^{k^3z}}dz + \frac{b}{2\delta}\int_{\mathcal{B}}\abs{\theta_kBe^{-k^3z}}dz
    \\ \nonumber
    %\le \frac{b\norm{\theta_k'}_2}{2} \Bigg[\abs{A}\sqrt{\frac{1}{2}k^{-3}(e^{2k^3}-e^{2(1-\delta)k^3}+e^{2\delta k^3}-1)} + \abs{B}\sqrt{\frac{1}{2}k^{-3}(1-e^{-2\delta k^3}+e^{-2(1-\delta)k^3}-e^{-2k^3})}\Bigg] 
    %\\ \nonumber
    &= \frac{\sqrt{2}b\norm{\theta_k'}_2}{4k^{\frac{-3}{2}}}(\abs{A}e^{k^3}+\abs{B})\sqrt{1-e^{-2\delta k^3}+e^{-2(1-\delta)k^3}-e^{-2k^3}}
    \\ \nonumber
    &\le \frac{b\Ray\norm{\theta_k'}_2\norm{\theta_k}_2}{2k^2}\sqrt{1-e^{-2\delta k^3}+e^{-2(1-\delta)k^3}-e^{-2k^3}}
    \\ \nonumber
    &= \frac{b\Ray\norm{\theta_k'}_2\norm{\theta_k}_2}{2k^2} \sqrt{(1-e^{-2\delta k^3})(1+e^{-2(1-\delta)k^3})}
    \\ \nonumber 
    &\le \frac{b\Ray\norm{\theta_k'}_2\norm{\theta_k}_2}{2k^2}2\sqrt{\delta}k^{\frac{3}{2}} 
    \\
    &= \frac{b\Ray\norm{\theta_k'}_2\norm{\theta_k}_2}{\sqrt{k}}.
    \label{eqn:BL_largeK}
\end{align}
The last inequality follows from noting that $e^{-2\delta k^3} \ge 1-2\delta k^3$, so $(1-e^{-2\delta k^3})(1+e^{-2(1-\delta)k^3}) \le 2\delta k^3(1+e^{-2(1-\delta) k^3}) \le 4\delta k^3$.
Note that \eqref{eqn:BL_largeK} is valid for all $k > 1$, but is most valuable for $k \gg 1$.

%%%%%%%%%%%%%%%%%%%%%%%%%%%%%%%%%%
\subsection{Contribution of exponential terms for $k\leq 1$}\label{s:bl-estimates-small-k}
As before, to handle small wavenumbers we rewrite $Ae^{k^3z} + Be^{-k^3z} = c_1\sinh{k^3z} + c_2\cosh{k^3z}$. Following a similar approach as in the previous section, we find
\begin{equation}
    \begin{split}
\left|    \frac{b}{2\delta}\int_{\mathcal{B}}\theta_k(Ae^{k^3z}+Be^{-k^3z})dz\right| \le
    \frac{b\norm{\theta_k'}_2}{2} \Bigg[\abs{c_1}\sqrt{\frac{1}{4}k^{-3}(\sinh{2k^3}-\sinh{\left[2(1-\delta)k^3\right]}+\sinh{2\delta k^3}-4\delta k^3)} \\+ \abs{c_2}\sqrt{\frac{1}{4}k^{-3}(\sinh{2k^3}-\sinh{\left[2(1-\delta)k^3\right]}+\sinh{2\delta k^3}+4\delta k^3)}\Bigg].
    \end{split}
\end{equation}
Noting that $\sinh(b) - \sinh(a) = \int_a^b \cosh{x} dx$ and $\cosh{x}$ is increasing for $x > 0$, we recognize that $\sinh(b) - \sinh(a) \le (b-a)\cosh{b}$ provided $0 \leq a \leq b$.  Setting $a=0,~b=2\delta k^3$ and $a=2(1-\delta)k^3,~b=2k^3$, respectively, we can estimate the two terms on the right-hand side of the last bound to obtain
\begin{multline}
    \left|\frac{b}{2\delta}\int_{\mathcal{B}}\theta_k(Ae^{k^3z}+Be^{-k^3z})dz\right|
        \le \frac{b\norm{\theta_k'}_2}{2} \Bigg[\abs{c_1}\sqrt{\frac{\delta}{2}(\cosh{2k^3}+\cosh{2\delta k^3}-2)} \\ + \abs{c_2}\sqrt{\frac{\delta}{2}(\cosh{2k^3}+\cosh{2\delta k^3}+2)}\Bigg].
\end{multline}
Since $\cosh{x} \le 1 + x^2$ whenever $0 \le x \le 2$ and since we are considering $k \le 1$, we can further estimate
\begin{align}
        \left|\frac{b}{2\delta}\int_{\mathcal{B}}\theta_k(Ae^{k^3z}+Be^{-k^3z})dz\right| 
        &\le \frac{b\sqrt{\delta}\norm{\theta_k'}_2}{2} \Bigg[\abs{c_1}\sqrt{2k^6+2\delta^2 k^6} + \abs{c_2}\sqrt{2k^6+2\delta^2 k^6+2}\Bigg]
        \nonumber \\
        & \le \frac{b\sqrt{\delta}\Ray\norm{\theta_k'}_2\norm{\theta_k}_2}{4} \Bigg[2k^{\frac{5}{2}}\sqrt{1+\delta^2} + \sqrt{2\epsilon k (1 + k^6(1+\delta^2))}\Bigg]
        \nonumber \\
        &\le \frac{b}{2}\sqrt{\delta + \delta^3}\Ray\norm{\theta_k'}_2\norm{\theta_k}_2\Bigg[k^{\frac{5}{2}} + \sqrt{\epsilon k}\Bigg],
\label{eqn:BL_smallK}
\end{align}
where the last line above relies on the restriction to $k\leq 1$ and the obvious inequality $1 \leq 1+\delta^2$.

%%%%%%%%%%%%%%%%%%%%%%%%%%%%%%%%%%%%%%%%%%%%
\subsection{An estimate for all wavenumbers}\label{s:matching-bounds}
In the last two subsections we have established bounds of the form
\begin{equation}
    \frac{b}{2\delta}\int_{\mathcal{B}}\abs{\theta_k(Ae^{k^3z}+Be^{-k^3z})dz} \leq p(k) \norm{\theta_k'}_2\norm{\theta_k}_2
\end{equation}
for $k > 1$  and $k\leq 1$ separately. We now turn to the problem of bounding
\begin{equation}
\label{e:bound-exponential-term-uniform-k}
    \frac{b}{2\delta}\int_{\mathcal{B}}\abs{\theta_k(Ae^{k^3z}+Be^{-k^3z})dz} \leq d_1(k^2\norm{\theta_k}^2_2 + \epsilon^2\norm{\theta_k'}^2_2)
\end{equation}
for some positive constant $d_1$ uniformly in $k$, which is what's needed to control the sign-indefinite integrals in~\eqref{e:analytical-Sk} using the positive definite terms. 

On the one hand, by Young's inequality we have 
\begin{equation}
    \label{e:d1-young}
    d_1(k^2\norm{\theta_k}^2_2 + \epsilon^2\norm{\theta_k'}^2_2) \ge 2 d_1 k\epsilon\norm{\theta_k'}_2\norm{\theta_k}_2.
\end{equation} 
On the other hand, the estimate $\norm{\theta_k}_2 \le \frac{1}{2}\norm{\theta_k'}_2$ implies
\begin{equation}
    d_1(k^2\norm{\theta_k}^2_2 + \epsilon^2\norm{\theta_k'}^2_2) \ge
    d_1\epsilon^2\norm{\theta_k'}^2_2 \ge 2 d_1\epsilon^2\norm{\theta_k}^2\norm{\theta_k'}^2.
\end{equation}
Consequently, the desired bound \eqref{e:bound-exponential-term-uniform-k} holds if
\begin{equation}
    p(k) \le \max(2d_1\epsilon^2, 2d_1k \epsilon).
\end{equation}

Combining the bounds for small and large $k$ in \eqref{eqn:BL_largeK} and \eqref{eqn:BL_smallK} we find
\begin{equation}
    p(k)  = \begin{cases}\frac{b}{2}\Ray \sqrt{\delta + \delta^3}(k^{\frac{5}{2}} + \sqrt{\epsilon k}) & k \le 1\\
    b\Ray \sqrt{\frac{\delta}{k}} & k > 1.
    \end{cases}
\end{equation}
If $k > 1 > \epsilon$, then the bound holds with $p(k) = b\Ray\sqrt{\frac{\delta}{k}} \le 2d_1\epsilon k$, so $\sqrt{\frac{\delta}{k}} \le \frac{2d_1k\epsilon}{b\Ray}$. Hence, the bound holds for $k > 1$ if
\begin{equation}
 \delta \le \frac{4d_1^2\epsilon^2k^3}{b^2\Ray^2}.
\end{equation}
This must hold for all $k$ > 1, which in the limit of $k \to 1^+$ yields
\begin{equation}\label{eq:delta_largeK}
\delta \le \frac{4d_1^2\epsilon^2}{b^2\Ray^2}.
\end{equation}

For $k \in [\epsilon, 1]$, instead, we require 
\begin{equation}
    \sqrt{\delta + \delta^3}\left(k^{\frac{5}{2}} + \sqrt{\epsilon k}\right) \le \frac{4d_1k\epsilon}{b\Ray},
\end{equation}
or, equivalently,
\begin{equation}
    \sqrt{\delta + \delta^3}\left(k^{\frac{3}{2}} + \sqrt{\frac{\epsilon}{k}}\right) \le \frac{4d_1\epsilon}{b\Ray}.
\end{equation}
The left-hand side of the last inequality is a convex function of $k$, hence maxima can only occur on the boundaries of the interval $[\epsilon, 1]$. Thus, we require
\begin{equation}
    \sqrt{\delta + \delta^3}\max(1+\sqrt{\epsilon}, 1 + \epsilon^{\frac{3}{2}}) = \sqrt{\delta + \delta^3}(1+\sqrt{\epsilon}) \le \frac{4d_1\epsilon}{b\Ray}.
\end{equation}
Recall that $\delta < \frac{1}{2}$ and $\epsilon < 1$, implying that
\begin{equation}
\sqrt{1+\delta^2}(1+\sqrt{\epsilon}) \le \sqrt{\frac{5}{4}}2 \le \sqrt{5}.
\end{equation}
Hence, if
\begin{equation}
\sqrt{\delta} \le \frac{4d_1\epsilon}{\sqrt{5}b\Ray} \le \frac{4d_1\epsilon}{\sqrt{1+\delta^2}(1+\sqrt{\epsilon})b\Ray},
\end{equation}
the desired bound holds for $k \le 1$. We therefore require that
\begin{equation}
    \delta \le \frac{16d_1^2\epsilon^2}{5b^2\Ray^2} \le \frac{4d_1^2\epsilon^2}{b^2\Ray^2},
\end{equation}
which is the same as~\eqref{eq:delta_largeK}. 

Finally, if $k < \epsilon$, then we require $p(k) \le 2d_1\epsilon^2$. This reduces to 
\begin{equation}
    \sqrt{\delta + \delta^3}(k^{\frac{5}{2}} + \sqrt{\epsilon k}) \le \frac{4d_1\epsilon^2}{b \Ray}.
\end{equation}
The left-hand side of this inequality is clearly an increasing function in $k$, while the right-hand side is constant. Taking the limit $k \to \epsilon$, reduces the problem to the previous case, so the last inequality is satisfied whenever \eqref{eq:delta_largeK} holds. In summary, condition \eqref{eq:delta_largeK} ensures that \eqref{e:bound-exponential-term-uniform-k} holds uniformly in~$k$.

%%%%%%%%%%%%%%%%%%%%%%%%
\subsection{Boundary integral of $\theta_k h_k$}\label{s:bl-estimates-green}
We now estimate the contribution to \eqref{e:indefinite-terms} of the boundary integral of $\theta_k h_k$. Precisely, we seek to bound 
\begin{equation}
    \label{e:estimate-green}
    \frac{b}{2\delta}\int_{\mathcal{B}}\abs{h_k(z)\theta_k(z)}dz \le d_2(k^2\norm{\theta_k}^2_2 + \epsilon^2\norm{\theta_k'}^2_2)
\end{equation}
for some constant $d_2 > 0$ to be specified later. 

Recall that
\begin{equation}
    h_k(z) = \frac{\Ray k}{\sinh{k^3}}\int^{1}_{0} g(z,s)\theta_k(s)ds,
\end{equation}
  where 
  \begin{equation}
      g(z,s) = \begin{cases} 
      \sinh{k^3z}\sinh{k^3(1-s)} & z \leq s \\
      \sinh{k^3s}\sinh{k^3(1-z)} & s \leq z 
   \end{cases}.
\end{equation}
The analysis in \cite{Grooms_2014} yields %the following piece-wise bound bound on the integral of $h_k$ over the boundary layers:
\begin{equation}
    \left(\int_0^\delta h_k(z)^2dz\right)^{\frac{1}{2}} \le
    \begin{cases}
        \delta\Ray\norm{\theta_k}_2\sqrt{\delta k^5 P} & k^3\delta \le \gamma \\
        \frac{\delta}{2}\Ray\norm{\theta_k}_2\sqrt{\frac{L}{2k\delta}} & k^3\delta \ge \gamma
    \end{cases},
\end{equation}
where $\sinh{\gamma} = 1$, $P$ = $2(\sinh{1}-1)$, and 
\begin{equation}
\label{e:L-def}
L = \bigg(\frac{\gamma}{2}+4\coth{2\gamma}+\frac{\csch^2(2\gamma)}{\gamma}\bigg).
\end{equation}
The same piece-wise bound also holds when the integration interval $[0,\delta]$ is replaced by $[1-\delta,1]$. Hence,
\begin{align}\nonumber
    \frac{b}{2\delta}\int_{\mathcal{B}}\abs{h_k(z)\theta_k(z)}dz 
    &\le \frac{b\norm{\theta_k'}}{2} \left(\int_B h_k^2dz\right)^{\frac{1}{2}}\\ \nonumber
    &= \frac{b\norm{\theta_k'}}{2} \left(\int_0^\delta h_k^2dz + \int_{1-\delta}^{1}h_k^2\right)^{\frac{1}{2}}\\ \nonumber
    &\le \frac{b\norm{\theta_k'}}{2} \Bigg[2\max\left(\int_0^\delta h_k^2dz, \int_{1-\delta}^{1}h_k^2\right)\Bigg]^{\frac{1}{2}}\\
    &\le 
    \begin{cases}
        \frac{\sqrt{2}}{2}b\delta\Ray\norm{\theta_k}_2\norm{\theta_k'}_2\sqrt{\delta k^5 P} & k^3\delta \le \gamma \\
        \frac{\sqrt{2}}{4}b\delta\Ray\norm{\theta_k}_2\norm{\theta_k'}_2\sqrt{\frac{L}{2k\delta}} & k^3\delta \ge \gamma
    \end{cases}.
\end{align}

We first consider the case $k^3\delta \ge \gamma$. Arguing as for~\eqref{e:d1-young}, the bound~\eqref{e:estimate-green} holds for any fixed $d_2$ provided that
\begin{equation}
\frac{\sqrt{2}}{4}b\delta \sqrt{\frac{L}{2k\delta}}\Ray \le 2d_2k\epsilon,
\end{equation}
which is equivalent to
\begin{equation}
\sqrt{\delta} \le \frac{8d_2k^{\frac{3}{2}}\epsilon}{b\Ray\sqrt{L}}.
\end{equation}
Squaring both sides and taking the minimum over admissible $k$, which is attained for $k^3$=$\gamma/\delta$, we conclude that the last inequality holds if
\begin{equation}\label{eq:delta_hk1}
    %\delta \le \frac{64d_2^2\gamma\epsilon^2}{b^2\Ray^2L\delta} \implies 
    \delta \le \frac{8d_2\sqrt{\gamma}\epsilon}{b\Ray\sqrt{L}}.
\end{equation}

Similarly, for $k^3\delta < \gamma$ we require
\begin{equation}
\frac{\sqrt{2}}{2}b\delta\Ray\sqrt{\delta k^5P} \le 2d_2k\epsilon,
\end{equation}
which holds if
\begin{equation}
\frac{\sqrt{2}}{2}b\delta\Ray\sqrt{\delta k^3P} \le 2d_2\epsilon.
\end{equation}
Upon maximizing the left-hand side over $k$ we arrive at the condition
\begin{equation}\label{eq:delta_hk2}
    \delta \le \frac{2\sqrt{2}d_2\epsilon}{b\Ray\sqrt{\gamma P}}
\end{equation}
Since $8\sqrt{\frac{\gamma}{L}} \approx 3.4 \le 5.1 \approx \frac{2\sqrt{2}}{\sqrt{\gamma P}}$ we conclude that \eqref{eq:delta_hk2} is automatically satisfied whenever \eqref{eq:delta_hk1} holds, so we only consider the latter in what follows.

%%%%%%%%%%%%%%%%%%%%
\subsection{The final bound}\label{s:final-bound}
We now collect all estimates from the previous sections to optimize the parameters $\delta$, $b$, $d_1$ and $d_2$ such that \eqref{e:analytical-Sk} holds for all wavenumbers $k$ and the bound \eqref{e:analytical-bound-formula} on the Nusselt number is as small as possible.

We begin with the observation that, upon setting
\begin{equation}
\label{e:optimal-delta-min}
\delta = \min\left(\frac{8d_2\sqrt{\gamma}\epsilon}{b\Ray\sqrt{L}},\; \frac{16d_1^2\epsilon^2}{5b^2\Ray^2}\right),
\end{equation}
we have 
\begin{align}
    \frac{b}{2\delta}\int_{\mathcal{B}}\abs{\theta_kw_k}dz
    &\le  \frac{b}{2\delta}\int_{\mathcal{B}}\abs{\theta_kh_k}dz + \frac{b}{2\delta}\int_{\mathcal{B}}\abs{\theta_k(Ae^{k^3z}+Be^{-k^3z})}dz
    \nonumber \\
    &\le (d_1+d_2)(k^2\norm{\theta_k}^2_2 + \epsilon^2\norm{\theta_k'}^2_2).
\end{align}
Thus, \eqref{e:analytical-Sk} holds for all $k$ provided $d_1+d_2 \le b-1$. 

To optimize $d_1$ and $d_2$, therefore determine $\delta$, observe from \eqref{e:analytical-bound-formula} that to minimize the eventual bound on $\Nu$ 
it is desirable to maximize the boundary layer width $\delta$. This, in turn, requires $d_1$ and $d_2$ to be as large as possible. Thus, we set $d_2 = b - 1 - d_1$ and determine $d_1$ by requiring that
\begin{equation}
\frac{8d_2\sqrt{\gamma}\epsilon}{b\Ray\sqrt{L}} = \frac{16d_1^2\epsilon^2}{5b^2\Ray^2},
\end{equation}
so the two arguments of the $\min$ in~\eqref{e:optimal-delta-min} match. After some simple algebra one finds that the optimal $d_1$ solves the quadratic equation
\begin{equation}
    %d_1^2 - \frac{5b\sqrt{\gamma}\Ray}{2\sqrt{L}\epsilon}(b-1-d_1) = 0 \implies 
    d_1^2 + Md_1 - (b-1)M = 0,
\end{equation}
where
\begin{equation}
M := \frac{5b\sqrt{\gamma}\Ray}{2\sqrt{L}\epsilon}.
\end{equation}
Choosing the positive solution we arrive at
\begin{align}
        d_1 &= \frac{1}{2}\left(\sqrt{M^2+4(b-1)M} - M\right) \nonumber\\
        &= \frac{4(b-1)M}{2(\sqrt{M^2+4(b-1)M} + M)} \nonumber \\
        &= \frac{2(b-1)}{\sqrt{1 + \frac{4(b-1)}{M}} + 1},
\end{align}
which can be substituted into \eqref{e:optimal-delta-min} to obtain the optimal $\delta$, i.e.:
\begin{equation}
    \delta = \frac{16 d_1^2\epsilon^2}{5b^2\Ray^2} = \frac{64(b-1)^2}{5b^2\left(2 + \frac{4(b-1)}{M} + 2\sqrt{1+\frac{4(b-1)}{M}}\right)}\frac{\epsilon^2}{\Ray^2} \sim \frac{64(b-1)^2}{10b^2}\frac{\epsilon^2}{\Ray^2},
\end{equation}
where the final approximation holds for large values of $\Ray$ and/or small values of $\epsilon$.  Substituting this value into \eqref{e:analytical-bound-formula} and optimizing the coefficient of the term proportional to $\Ray^2 \epsilon^{-2}$ (the dominant term illustrated above) over $b$ gives the optimal choice $b=4$ and the complete upper bound
\begin{equation}
    \Nu \leq \frac{5\bigg(\sqrt{1 + \frac{12}{M}} + 1\bigg)^2\Ray^2}{54\epsilon^2} - \frac{1}{3}.
\end{equation}

To yield a more readable dependence of this bound on $\Ray$ and $\epsilon$ and obtain~\eqref{e:final-bound-Nu} observe that, for any positive $x$,
\begin{equation}
\sqrt{1+x}-1 = \frac{x}{\sqrt{1+x}+1} \le \frac{x}{2}.
\end{equation}
Using this estimate we find 
\begin{equation}
    \bigg(\sqrt{1 + \frac{12}{M}} + 1\bigg)^2 = 2 + \frac{12}{M} + 2\sqrt{1 + \frac{12}{M}} \le 4 + \frac{24}{M}.
\end{equation}
Now recalling that, for $b=4$,
\begin{equation}
M = \frac{5b\sqrt{\gamma}\Ray}{2\sqrt{L}\epsilon} = \frac{10\sqrt{\gamma}\Ray}{\sqrt{L}\epsilon},
\end{equation}
we have 
\begin{align}\nonumber
    \Nu 
    &\le \frac{10}{27}\Ray^2\epsilon^{-2} + \frac{20\Ray^2}{9\epsilon^2M} - \frac{1}{3}\\
    &=  \frac{10}{27}\Ray^2\epsilon^{-2} + \frac{2\sqrt{L}}{9\sqrt{\gamma}}\Ray\epsilon^{-1} - \frac{1}{3}.
\end{align}
We obtain~\eqref{e:final-bound-Nu} upon rewriting this in terms of the original Rayleigh and Ekman numbers using the identities $\Ray = \Ra \Ek^{4/3}$ and $\epsilon = \Ek^{1/3}$, and recalling the definition of $L$ from~\eqref{e:L-def}.

\section{Discussion and conclusions}\label{sec:conc}
We have applied the auxiliary functional method \cite{chernyshenko2014polynomial} to derive (modulo small corrections for $\Ra \Ek \gg 1$) the bound $\Nu\le 0.3704 \Ra^2\Ek^2$ for heat transport in rapidly-rotating Rayleigh-B\'enard convection at infinite Prandtl number between no-slip boundaries.
The bound applies to the asymptotically reduced model of \cite{julien2016nonlinear,plumley2016effects}, and updates an earlier bound for stress-free boundaries by \cite{Grooms_2014}.

Previous work on rotating Rayleigh-B\'enard convection at infinite Prandtl number has produced the following results for no-slip boundaries: $Nu\le 1 + 9.5\Ra^2\Ek$ from \cite{constantin1999heat}, $\Nu\le 0.6635\ldots Ra^{2/5}$ from \cite{doco2001}, and $\Nu\le c\Ra^{4/11}(1 + 1/(2\Ek))^{4/11}$ with $c<2$ from \cite{yan2004limits}.
These bounds were derived directly from the incompressible Navier-Stokes equations in the Boussinesq approximation, rather than from the asymptotically reduced equations.
In the limit of rapid rotation (small $\Ek$), our bound is already a factor of $\Ek$ better than the next-best bound from \cite{constantin1999heat}, and is in fact comparable to a bound for stress-free boundaries derived in the same paper. Moreover, the apparent disparity between our rigorous upper bound and the bounds optimized numerically shown in figure~\ref{fig:optimized-bounds} suggests that even better bounds may be proved. Unfortunately, however, the range of parameters covered by the numerical results of section \ref{s:af-method} is not large enough to anticipate the optimal scaling of upper bounds obtained with quadratic auxiliary functions. This limitation is due to the large number of Legendre expansion coefficients required to capture very steep boundary layers in $\phi(z)$ (cf. figure~\ref{f:optimal-profiles}) and the numerical challenges associated with setting $\varepsilon \ll 1$ in~\eqref{eq:Sk-inequality}.

Current phenomenological and empirical scaling laws for the heat transport in rapidly-rotating Rayleigh-B\'enard convection span the gamut from $\Nu\sim \Ra \Ek^{4/3}$ in \cite{ecke2014heat} to $\Nu\sim \Ra^{3.6} \Ek^{4.8}$ in some experiments of \cite{cheng2015laboratory}, though it bears noting that these are developed at finite Prandtl numbers.
Since the onset of convection occurs for $\Ra\sim \Ek^{4/3}$, it seems natural to expect the Nusselt number to scale as $\Nu\sim(\Ra \Ek^{4/3})^\alpha$ for some range of Rayleigh numbers near onset and some unknown $\alpha$.
For stress-free boundaries it has been predicted that an initial steep $\alpha$ eventually gives way to $\alpha=3/2$ as the Rayleigh number increases \cite{julien2012statistical}, which corresponds to behavior predicted by an asymptotic analysis of exact steady laminar solutions in \cite{grooms2015asymptotic}.
Numerical computations with exact laminar solutions of the asymptotically reduced model with no-slip boundaries \cite{julien2016nonlinear} and turbulent simulations of the reduced model \cite{plumley2016effects,plumley2017sensitivity} show the heat transport scaling exponent $\alpha$ behaving similarly: it diminishes to $\alpha=3/2$ as the Rayleigh number increases.
However, the range of Rayleigh numbers over which the initial large values $\alpha$ hold increases with the Prandtl number, and simulations have not yet examined the infinite Prandtl number case.
Moreover, the transition from a large $\alpha$ to $\alpha=3/2$ has yet to be observed in laboratory experiments: Cheng et al. saw the exponent $\alpha$ increase with decreasing $\Ek$ up to a value of $\alpha=3.6$ with no signs of a transition back to $\alpha=3/2$.

Writing our bound as $\Nu\le0.3704(\Ra\Ek^{4/3})^2\Ek^{-2/3}$ shows that at fixed Ekman number our exponent is slightly higher than the currently hypothesized inviscid scaling law of $\alpha=3/2$ from \cite{julien2012statistical}.
This discrepancy could perhaps be explained by noting that an inviscid scaling is hardly appropriate to an infinite Prandtl regime, and that the heat transport in simulations tends to increase with increasing Prandtl number.
At the same time, our bound is consistent with the behavior observed in \cite{cheng2015laboratory}, where heat transport increases more and more rapidly with decreasing $\Ek$.
Although the exponent on the Rayleigh number in our bound remains at 2 regardless of the Ekman number, the prefactor to $(\Ra \Ek^{3/4})^2$ increases without bound as $\Ek\to0$.

\begin{acknowledgements}
JPW would like to thank K. Julien, M. Plumley, J. Aurnou, J. S. Cheng, and R. Kuhnnen for insightful discussions and feedback related to this problem. In particular, M. Plumley and J. S. Cheng provided the numerically comptued values that generate the comparison in figure \ref{fig:optimized-bounds}.  JPW also acknowledges support
from the Simon’s Foundation through award number 586788.  GF was supported in part by an EPSRC Doctoral Prize Fellowship and in part by an Imperial College Research Fellowship.
\end{acknowledgements}

%\nocite*{}
%\bibliography{aipsamp}%
\bibliography{references.bib}
%Produces the bibliography via BibTeX.

\end{document}